\documentclass[10pt,letterpaper]{article}
\usepackage{opex3}
\usepackage{amsmath}
\usepackage{amssymb}
\usepackage{bm}

\def\be{\begin{eqnarray}}
\def\ee{\end{eqnarray}}

\def\E{{\bf E}}

\def\B{{\bf B}}

\def\p{{\bf p}}
\def\m{{\bf m}}

\def\u{{\bf u}}

\begin{document}

\title{Strong magnetic response of submicron Silicon particles in the infrared}

\author{A. Garc\'{\i}a-Etxarri ,$^1$ R.  G\'omez-Medina,$^2$ L. S. Froufe-P\'erez,$^2$ \\ C. L\'opez,$^2$ L.
Chantada,$^3$  F. Scheffold,$^3$ J. Aizpurua,$^1$  \\ M. Nieto-Vesperinas $^2$ and J. J.
S\'aenz$^{1,4}$}

\address{$^1$ Donostia
International Physics Center (DIPC), Paseo Manuel Lardizabal 4, 20018
Donostia-San Sebastian, Spain}
\address{$^2$Instituto de Ciencia de  Materiales de Madrid, CSIC, Campus de
Cantoblanco, 28049 Madrid, Spain}
\address{$^3$ Department of Physics, University of Fribourg,  Chemin
du Musée 3, 1700 Fribourg, Switzerland}
\address{$^4$Departamento de F\'{\i}sica de la Materia Condensada, Universidad
Aut\'{o}noma de Madrid, 28049 Madrid, Spain}

\email{juanjo.saenz@uam.es}


\begin{abstract}
High-permittivity dielectric particles with resonant magnetic properties are being explored as
constitutive elements of new metamaterials  and devices in the microwave regime.
Magnetic properties of low-loss dielectric nanoparticles in the visible or infrared are not expected due
to intrinsic low refractive index of optical materials in these regimes.
Here we analyze the dipolar electric and magnetic response of  loss-less dielectric spheres made of moderate permittivity materials.
For low material refractive index ($\lesssim 3$) there are no sharp resonances due to
strong overlapping  between different multipole contributions. However,
we find that Silicon particles with  refractive index $\sim 3.5$ and radius $\sim 200$nm present a dipolar and strong magnetic
resonant response in telecom and near-infrared frequencies, (i.e. at
wavelengths $ \approx 1.2-2 \mu$m). Moreover, the light scattered by
these Si particles can be perfectly described by dipolar
electric and magnetic fields, quadrupolar and higher order
contributions  being negligible.
\end{abstract}

\ocis{(290.5850) Scattering, particles ; (160.1190)
Anisotropic optical materials ; (160.3820)   Magneto-optical materials.} 


\section{Introduction}

Electromagnetic scattering from nanometer-scale objects has long
been a topic of great interest and relevance to fields from
astrophysics or meteorology to biophysics and material science
\cite{1,2,3,4,Purcell,Draine}. During the last decade nano-optics
has developed itself as a very active field within the
nanotechnology community. Much of it has to do with plasmon
(propagating) based subwavelength optics and applications
\cite{Barnes}, the synthesis of negative-index optical metamaterials
\cite{Engheta_2005,Pendry,Shalaev} or the design of optical antennas
\cite{Antenna1,Antenna2,Niek}.

Magnetic effects, a key ingredient of relevant microwave
applications, cannot be easily exploited  in the optical range
(visible or infrared) due to intrinsic natural limitations of
optical materials. The quest for magnetic plasmons and magnetic
resonant structures at optical frequencies \cite{Engheta} has been
mainly focused on metallic structures with the unavoidable problems
of losses and saturation effects inherent to metals in the
optical range. Moreover, their size is often required to be
comparable to, or a fraction of, the operating wavelength in
order to provide a non-negligible response, and their magnetic
response is strongly affected and influenced by an intrinsic
quadrupolar contribution, which introduces additional radiation
losses and impurity of radiation and polarization \cite{10}.

As an alternative approach, there is a growing interest in the
theoretical and experimental study of high-permittivity dielectric
objects as constitutive elements of new metamaterials
\cite{Kevin_8,Wheeler_2009,Kevin_10,Kevin_11,Kevin_12,Kevin} and
antennas based on dielectric resonators
\cite{Mongia,Hsu,Brongersma_2009}. Interesting magnetodielectric
properties are usually associated to low loss and  large dielectric
constants which, being accessible  at  microwave and terahertz
frequencies,  are still a challenging issue in the infrared (IR) and
visible frequency ranges \cite{Review}. However, it has recently
been shown \cite{Kevin} that silicon rod arrays could exhibit a true
metamaterial left handed dispersion branch in the visible to mid-IR
despite the moderate value of the refractive index ($m \sim 3.5$).

Motivated by these results, we have analyzed in detail the
magnetodielectric properties of loss-less dielectric spheres made of
moderate permittivity materials in the infrared regime. Although the
analysis of scattering resonances of homogeneous spheres have been
analyzed in the past \cite{1,2,3}, we concentrate here on the
relevance of the first magnetic dipolar resonance in non-absorbing
dielectric nano-spheres (e.g. silicon spheres in the mid-IR). As we
will see, submicron Silicon spheres (with radius $a \approx 200$nm) can
present a strong magnetic resonance response at telecom and infrared
frequencies, (namely, wavelengths $\lambda \approx 2 n a \approx 1.5
\mu$m). Moreover, based on standard Mie theory \cite{1}, we have
found that {\it the scattering of these Si nanoparticles can be
perfectly described by dipolar electric and magnetic fields;
quadrupolar and higher order contributions being negligible}.
Interestingly, the magnetic resonance line remains  well defined and
resolved even for lower index materials, as Rutile-like TiO$_2$ with
an average index  $m \approx 2.8$ in the near-IR, even though in
this latter case the electric dipole resonance line strongly
overlaps with the magnetic dipole one, as well as with those of
higher order multipoles.

\section{Extinction resonances of a dielectric sphere. The Mie theory revisited}

\begin{figure}
\begin{center}
\includegraphics[width=6cm]{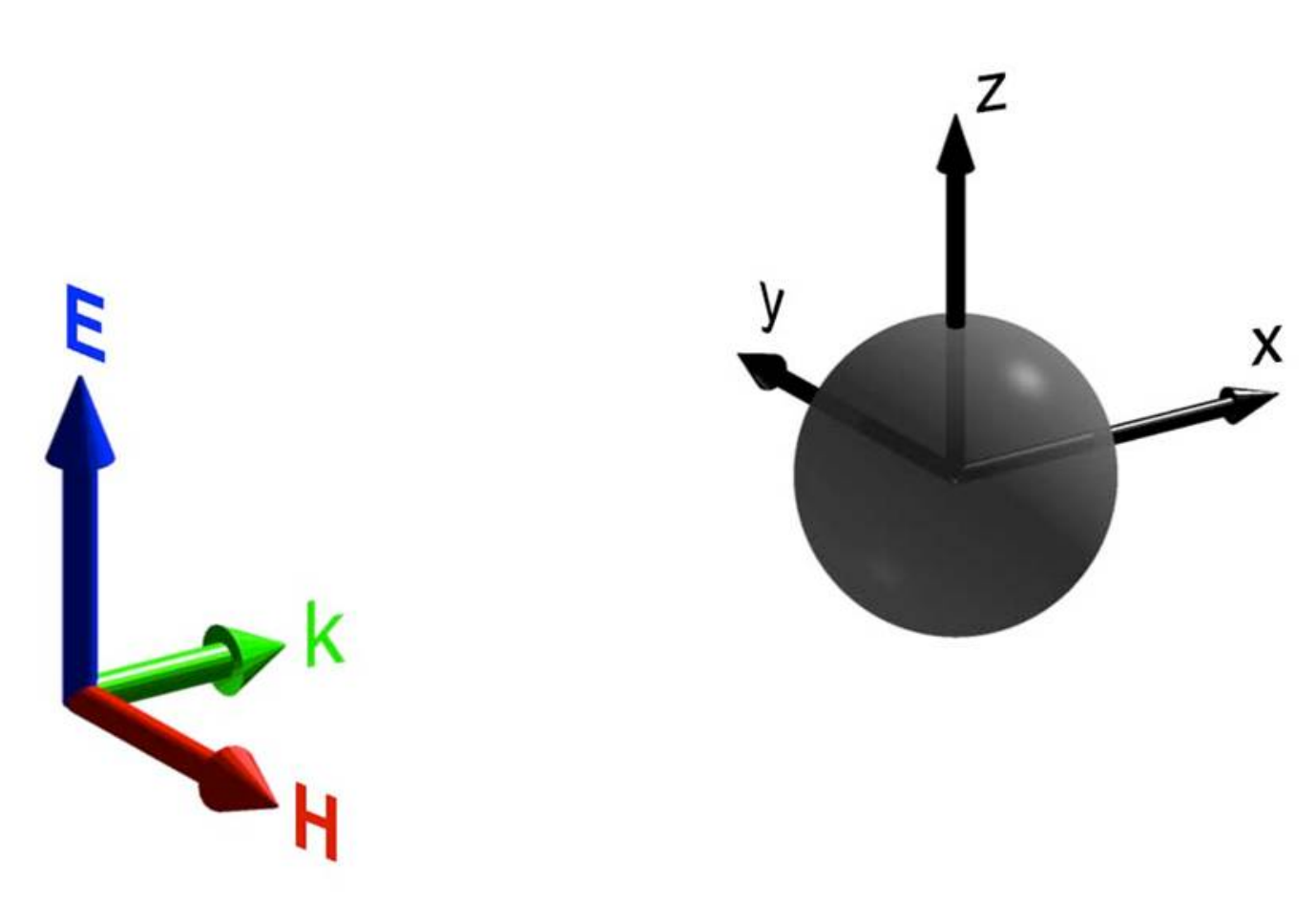}
\caption{Incident field vector}
\label{Sketch}
\end{center}
\end{figure}

Consider a non-absorbing dielectric sphere of radius $a$, index of
refraction $m_p$ and dielectric permittivity $\epsilon_p=m_p^2$ in
an otherwise uniform medium with real relative permittivity
$\epsilon_h$ and refractive index $m_h=\sqrt{\epsilon_h}$. The
magnetic permittivity of the sphere and the surrounding  medium is
assumed to be 1. Under plane wave illumination, and assuming
linearly polarized light, the incident wave is described by
 \be
 \E = E_0 \u_Z e^{ikX} e^{-i\omega t} \quad, \quad \B = B_0 \u_Y e^{ikX} e^{-i\omega t}
 \ee
 where  $k = m_h \omega/c =  m_h 2\pi/\lambda$, $\lambda$ being the wavelength in vacuum and $B_0 = \mu_0 H_0 =-(m_h/c) E_0 $ (see Fig. \ref{Sketch}). The
field scattered by the sphere can be decomposed into a multipole
series, (the so-called Mie's expansion), characterized by the
electric and  magnetic Mie coefficients  $\{a_n\}$ and $\{ b_n\}$,
respectively; ($a_1$ and $b_1$ being proportional to the electric
and magnetic dipoles, $a_2$ and $b_2$ to the quadrupoles, and so
on). We shall find useful to write the Mie coefficients in terms of
the scattering phase-shifts $\alpha_n$ and $\beta_n$ \cite{1} \be
a_n &=& \frac{1}{2}\left(1-e^{-2i\alpha_n }\right) = i \sin \alpha_n e^{-i\alpha_n} \label{an}\\
b_n &=& \frac{1}{2}\left(1-e^{-2i\beta_n }\right) = i \sin \beta_n e^{-i\beta_n} \label{bn}
\ee
where
\be
\tan \alpha_n &=&  \frac{m^2 j_n(y)\left[x j_n(x)\right]'- j_n(x)\left[y j_n(y)\right]'}
{ m^2 j_n(y) \left[x y_n(x)\right]' - y_n(x)\left[y j_n(y)\right]'}, \label{alphan} \\
\tan \beta_n &=&  \frac{ j_n(y)\left[x j_n(x)\right]'- j_n(x)\left[y
j_n(y)\right]'} { j_n(y) \left[x y_n(x)\right]' - y_n(x)\left[y
j_n(y)\right]'}, \label{betan} \ee
 $m=m_p/m_h$ being the relative
refractive index, $x = ka $  the size parameter and $y=mx$. $j_n(x)$
and $y_n(x)$ stands for the spherical Bessel and Neumann functions,
respectively, and the primes indicate differentiation with respect
to the argument. In absence of absorption, i.e. for real $m$, the phase
angles $\alpha_n$ and $\beta_n$ are real, then the extinction and
scattering  cross sections, $\sigma_{\text{ext}}$ and $\sigma_S$,
respectively, have the common value \cite{1}
\be \sigma_S =
\sigma_{\text{ext}} = \frac{2\pi}{k^2}\sum_{n=1}^\infty
\left(2n+1\right) \left\{\sin^2 \alpha_n+\sin^2 \beta_n \right\} =
\sum_{n=1}^\infty \left\{\sigma_{E,n}+\sigma_{M,n} \right\} \ee

\begin{figure}
\begin{center}
\includegraphics[width=12cm]{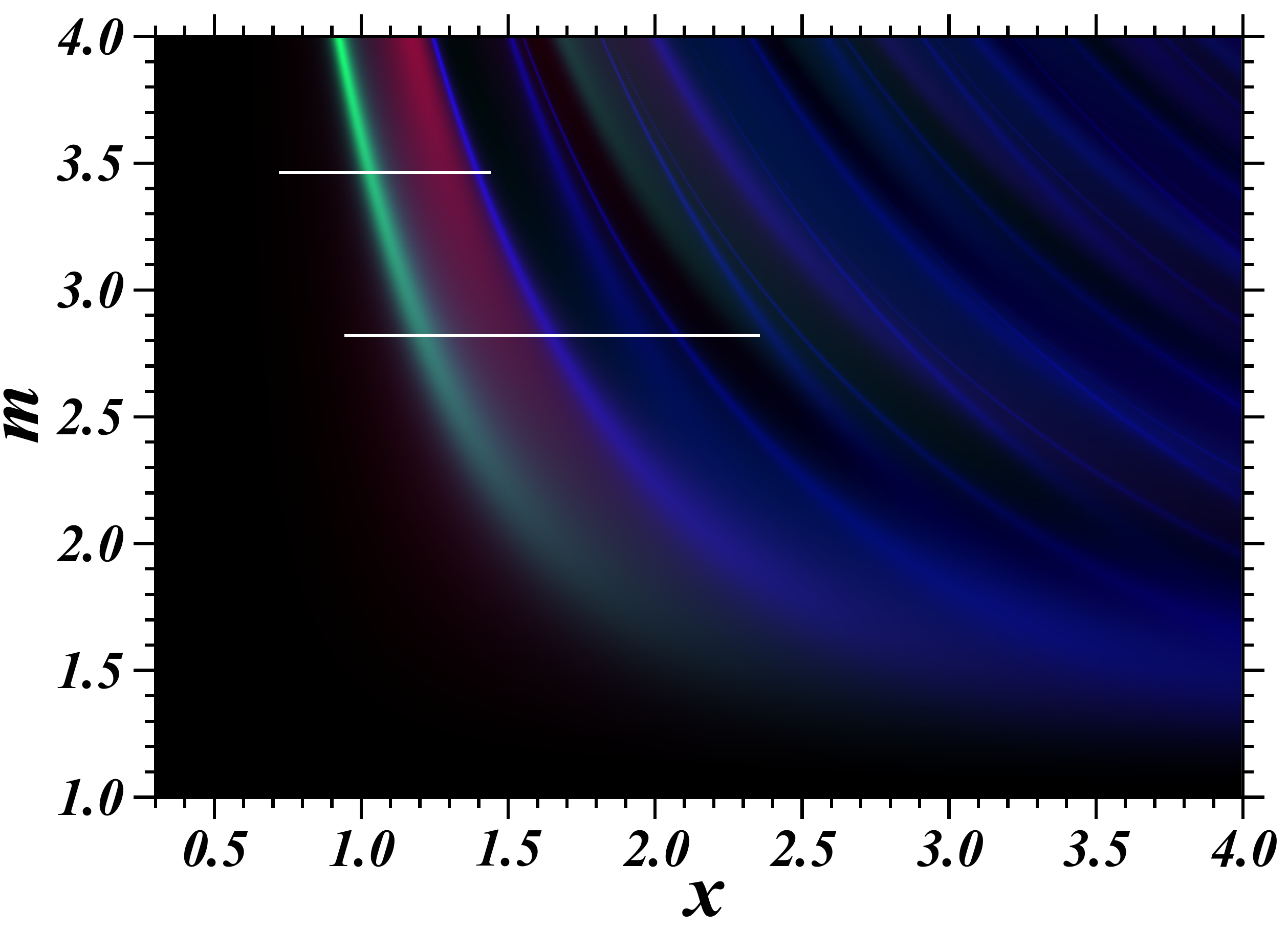}
\caption{Scattering cross section map of a non-absorbing Mie sphere as a function of the refractive
index $m$ and the size parameter $x= 2\pi a/\lambda$. Green
areas correspond to parameter ranges where the magnetic dipole contribution
dominates the total scattering cross section, while red areas represent
regions where the electric dipole contribution is dominating. The
remaining blue-saturated areas are dominated by higher order multipoles.
Brightness in the color-map is proportional to the total cross section.
White horizontal lines represent the $x$-range covered by figures \ref{SiMie} ($m =3.5$) and \ref{rutile} ($m= 2.8$)}.
\label{mapa}
\end{center}
\end{figure}

In the small particle limit ($x \ll 1$) and large particle
permittivities ($m \gg 1$) the extinction cross section presents
characteristic sharp resonance peaks. The values of  $y=mx$ at which
the angles $\alpha_n$ or $\beta_n$ are $\pi/2, 3\pi/2, \dots, etc$,
define the resonance points. At each resonance,  the extinction
cross section is of the order of $\lambda^2$ and it is independent
of either the particle size or the refractive index \cite{1}
 \be
 \sigma_S^{\text{res}} = \left.\sigma_{\text{ext}}^{\text{res}}\right|_{\{x \ll 1 \ ; \ m \gg 1\}} \approx \frac{2\pi}{k^2} (2n+1)
 \ee
 Asymptotically, the first resonance peak occurs at $y = \pi$ (i.e. $\lambda = m 2  a$) corresponding
 to the magnetic dipole term of coefficient $b_1$.

 Interesting consequences and applications of  well defined Mie resonance lines,
associated to low loss and  large dielectric constants, are
accessible for different materials  at  microwave and terahertz
frequencies. However, as $m$ decreases there is an increasing
overlap between the wavelength dependent cross-section peaks, and
the sphere resonant response smears out. Since usually non-absorbing
materials present low refractive index  in the infrared (IR) and
visible frequency ranges, Mie resonances of small particles in these
  regimes have not been considered in detail.
 However, the above analysis on the cross-section, carried out for Si nanoparticles in the IR,
 indicate that the aforementioned asymptotic behavior can be extended to a relative refractive index as low as
 $m \approx 3$.

\begin{figure}
\begin{center}
\includegraphics[width=12cm]{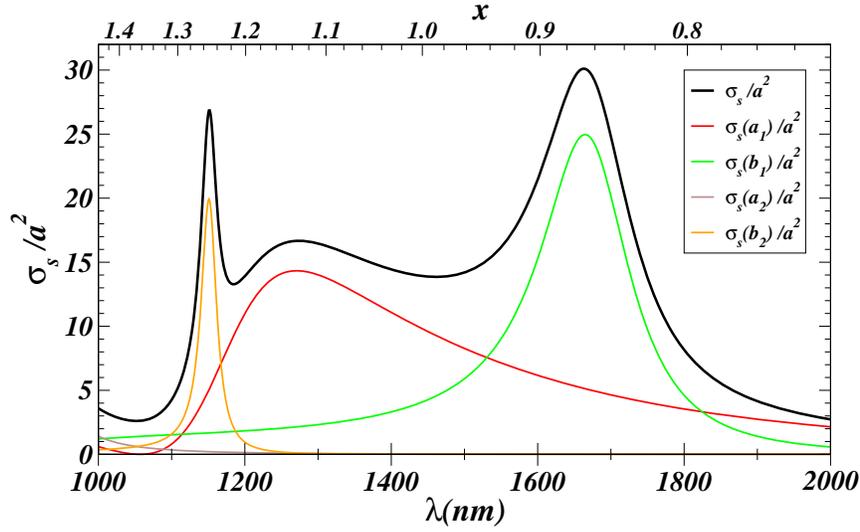}
\caption{Scattering cross-section $\sigma_S$ versus the wavelength $\lambda$ for a 230nm Si sphere (the refraction index $m=3.5$ is constant and real in this wavelength range). The contribution of each term in the Mie expansion is also shown. The green line corresponds to the magnetic dipole contribution.}
\label{SiMie}
\end{center}
\end{figure}

In Fig. \ref{mapa} we show the scattering cross section map of a non-absorbing Mie sphere as a function of the refractive
index $m$ and the size parameter $x= 2\pi a/\lambda$.
Brightness in the color-map is proportional to $\sigma_{\text{ext}}$ while color code represents
the contribution of electric and magnetic dipoles to the total cross
section. The RGB (Red, Green, Blue) code is formed by taking $RGB=\sigma_{E,1}R+\sigma_{M,1}G+\sigma_{\text{res}} B$
with $\sigma_{\text{res}}\equiv\sqrt{\sigma_{\text{ext}}^{2}-\sigma_{E,1}^2-\sigma_{M,1}}$.
The whole map is normalized to avoid over-saturation. Hence, green
areas correspond to parameter ranges where the magnetic dipole contribution
dominates the total scattering cross section, while red areas represent
regions where the electric dipole contribution is dominating. The
remaining blue-saturated areas are dominated by higher order multipoles.

\begin{figure}
\begin{center}
\includegraphics[width=10.5cm]{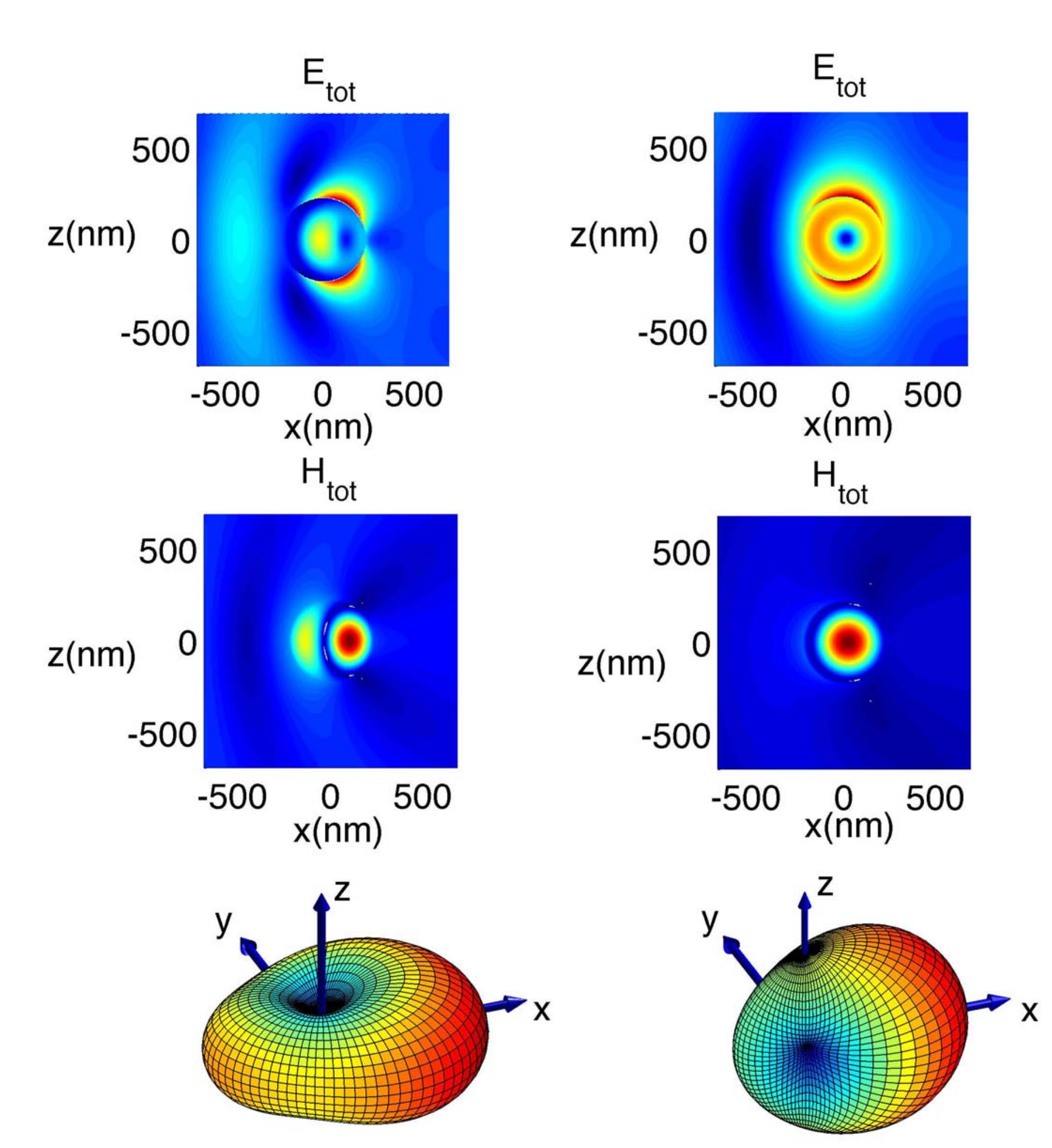}
\caption{Maps for the square modulus of the total electric an magnetic vectors, emitted by the  Si nanoparticle
of radius $a=230nm$  under plane wave illumination, (cf. Fig. \ref{Sketch}).
 The left and right panels correspond to the wavelengths $1250nm$ and $1680nm$ of the electric and magnetic resonance
peaks of Fig. \ref{SiMie}, respectively. The corresponding far field radiation patterns
are shown in the bottom. \label{Aitzol}
}
\end{center}
\end{figure}

In the micrometer wavelength regime, within the transparent region of
silicon ($\lambda \gtrsim 1100$ nm), the index of refraction can
well be approximated by a real constant $m_p \approx \sqrt{12} \sim
3.5$ (see for example \cite{Palik}). The scattering (or extinction)
cross section of a  Si sphere of radius $a=230nm$ in vacuum
($m_h=1$) is plotted in Fig. \ref{SiMie}. Although there is a
partial overlap between the first dipolar peaks, the magnetic line
of the first resonant peak (at $\lambda \approx 2 m a $) is still
very well resolved.
 Interestingly, for wavelengths larger than $\lambda \approx 1200$nm, the cross section is completely determined by the first $b_1$ and $a_1$ coefficients.
 In other words, in this regime Si particles can be treated as non-Rayleigh dipolar particles.

\section{Magnetic and electric resonances of dipolar particles}

Let us consider further in some detail the scattering properties of
small dielectric particles ($x\ll 1$). In the $x \ll 1$ limit, the
sphere is sufficiently small so that only the dipole scattered
fields are excited. The induced dipole moments, proportional to to
the external (polarizing) fields, $\E$ and  $\B$, are usually
written in terms of the particle electric and magnetic
polarizabilities $\alpha_E$ and $\alpha_M$, respectively : \be \p =
\epsilon_0 \epsilon_h \alpha_E \E \quad,\quad \m = \epsilon_0
\epsilon_h \alpha_M \B; \ee where
\begin{eqnarray}
 \alpha_E = i \left( \frac{k^3} {6\pi}\right)^{-1} a_1 \quad,\quad
\alpha_M = i \left( \frac{k^3} {6\pi}\right)^{-1} b_1 . \label{mMie}
\end{eqnarray}
By using the definition of the phase angles $\alpha_1$ and
$\beta_1$, [cf. Eqs. (\ref{an}) - (\ref{betan})], we can rewrite the
polarizabilities as
 \be
 \alpha_E = \frac{\alpha_E^{(0)}}{1-i\frac{k^3}{6\pi} \alpha_E^{(0)}} \quad, \quad \alpha_M =
 \frac{\alpha_M^{(0)}}{1-i\frac{k^3}{6\pi} \alpha_M^{(0)}}, \label{drainelike}
 \ee
where
 \be
 \alpha_E^{(0)} = - \frac{6\pi}{k^3} \tan \alpha_1 \quad, \quad \alpha_M^{(0)} = - \frac{6\pi}{k^3} \tan \beta_1
 \ee
 are the quasi-static polarizabilities. In terms of these magnitudes, the extinction and scattering cross sections
 then read
\be
\sigma_{\text{ext}} &=& k \text{Im}\left\{ \alpha_E + \alpha_M \right\} \\
\sigma_{S} &=&  \frac{k^4}{6\pi} \ \left\{ \left|\alpha_E\right|^2 +
\left|\alpha_M\right|^2 \right\} \ee Of course, in absence of
absorption, $\alpha_E^{(0)}$ and $\alpha_M^{(0)}$ are real
quantities and we recover the well known {\it optical theorem}
result $ \sigma_{\text{ext}} = \sigma_{S} $.

 In particular, in the Rayleigh limit, when $y \equiv 2\pi m a/\lambda \ll 1$, $\alpha_E^{(0)}$ and $\alpha_M^{(0)}$ approach the
 electrostatic form,
 \be
 \left.\alpha_E^{(0)}\right|_{y\ll 1} \approx  4\pi a^3 \frac{m^2-1}{m^2+2} \quad, \quad
 \left.\alpha_M^{(0)}\right|_{y\ll 1} \approx 4\pi a^3 (m^2-1) \frac{k^2a^2}{30}
\label{electrostatic} \ee From  Eq. (\ref{electrostatic}), together
with Eq. (\ref{drainelike}),  we recover the well known expression
for the polarizability of a Rayleigh particle including radiative
corrections \cite{Draine,Albadalejo_OE}. Notice that in this limit the magnetic
polarizability is negligible. Very small particles always behave as
point electric dipoles.

\begin{figure}
\begin{center}
\includegraphics[width=12cm]{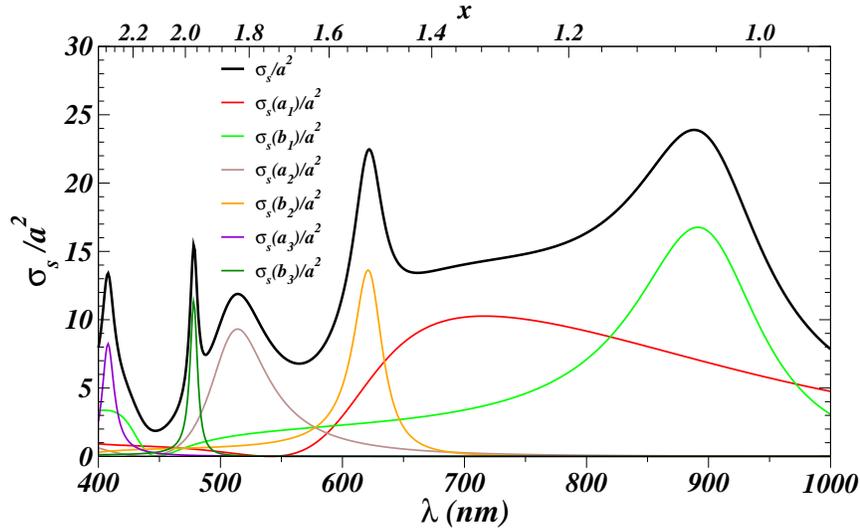}
\caption{Scattering cross-section $\sigma_S$ versus the wavelength $\lambda$ for a sphere of radius $a=150nm$
(its relative refractive index is $m=\sqrt{8}=2.82$, constant and real in this
wavelength range). The contribution of each coefficient in the Mie
expansion is also shown. The green line corresponds to the magnetic
dipole contribution.}
\label{rutile}
\end{center}
\end{figure}

 However, {\it as $y$ increases, (i.e. as $\lambda$ decreases), there is a crossover from electric to magnetic behavior}
 as shown by Fig. \ref{SiMie}
 for a Si particle. The $y$ values at which the quasi-static polarizability  $\alpha_E^{(0)}$ ( $\alpha_M^{(0)}$)
 diverges  define the electric (magnetic) dipolar resonances.
 {\it Near the first $b_1$-resonance, the particle  essentially behaves like a magnetic dipole}(cf. Fig. 2). If $\lambda$
 decreases further,  $a_1$ peak dominates and the sphere becomes again an electric dipole.  Notice however that,
 due to the overlap between the electric and magnetic responses, the radiation field near the resonances does not
 correspond to a fully pure electric or a fully pure magnetic dipolar excitation. Yet, as seen in Fig.2 for the
 aforementioned Si sphere,  the magnetic dipole contribution is, at its peak, about five times larger than the electric
 dipole one.

 As a further example, we have plotted in  Fig. \ref{Aitzol}
 both the near field intensity maps and far field radiation patterns at the resonant peaks of $a_1$ and $b_1$,
 (cf. Fig. \ref{SiMie}), numerically calculated from the full Mie solution (and not only from the dipolar approximation).
 The slightly distorted patterns with respect to the single dipole case are completely explained in terms of the sum
 of these  two mutually coherent electric  and magnetic dipoles.

 Silicon nanoparticles with radius of the order of 200-250nm are then fully characterized by  electric and magnetic
 polarizabilities as given by Eqs. \ref{drainelike} in the mid and near-IR region.
 By tuning the appropriate resonance wavelength, Si particles exhibit a strong magnetic
 response,  (only slightly perturbed by their electric dipole counterpart).

 It should be remarked that if the relative refractive index is smaller, the overlap between the Mie
 resonance lines  increases and then, a particular one single Mie term magnetic response is not well defined.
 As an example, in Fig. \ref{rutile}  we represent the scattering cross section of a 150nm radius particle with
 $m = \sqrt{8} \approx 2.8$
 (which would be of the order of the averaged refractive index of TiO$_2$ (rutile) in the near IR \cite{Palik}).
 As seen in this figure, although beyond 700nm the scattering is still described by the sum of the electric and magnetic
 dipoles,  the peak corresponding to the electric dipole, $a_1$, now disappears, and the overlapping between lines
 that produces this,  manifests a significant  contribution  of this electric
 dipole  at the magnetic resonance wavelength.

\section{Conclusion}

We have analyzed the dipolar electric and magnetic response of
lossless dielectric spheres made of both low and moderate
permittivity materials. Based on the Mie expansion,  we have derived
general expressions for the electric and magnetic polarizabilities
of dielectric spheres.  We found that submicron Silicon particles  present
a strong magnetic resonant  response in the mid-near infrared. Interestingly, {\it the light scattered by these Si
nanoparticles of appropriate size is perfectly described by dipolar
electric and magnetic fields, being quadrupolar and higher order
contributions negligible}. These results can play an important role,
not only in the field of metamaterials or optical antennas, but also
in tailoring the light transport through complex dielectric media
like photonic glasses \cite{Lopez,Frank} with intriguing magnetic
properties. Fabrication of new materials made of highly monodisperse
subwavelength silicon spheres \cite{Lopez2} may then lead to a new
generation of magnetodielectric optical materials. At the magnetic
or electric resonance wavelengths the extinction cross section is of
the order of $\lambda^2$  reaching its maximum theoretical limit
(independent of the particle size or refractive index) \cite{1} (
see also \cite{Brongersma_2009} ). The large dipolar cross-section
of magnetodielectric spheres near a magnetic resonance would also
imply strong radiation pressure magnetic forces
\cite{Albaladejo_prl,Chaumet_magnet,Nieto} leading to new concepts
related to the optical forces on magnetic particles .

\section*{Acknowledgement}
We appreciate interesting discussions with  S. Albaladejo, F. Gonz\'alez, L. Incl\'an, F. Moreno,  J.M. Saiz, I. Su\'arez and K. Vynck.
 This work has been
supported by the  EU  NMP3-SL-2008-214107-Nanomagma and NoE Nanophotonics4Energy 248855,  the
Spanish MICINN Consolider \textit{NanoLight} (CSD2007-00046), MAT2009-07841 GLUSFA and  FIS2009-13430-C02-02
and by the Comunidad de Madrid Microseres-CM (S2009/TIC-1476) and PHAMA-CM  (S2009/MAT-1756).
F.S. and L.C. acknowledge financial support from the Swiss National Science Foundation, Projects No. 126772 and 117762.
Work by R.G.-M. and L.S.F.-P. was supported by the MICINN ``Juan de la Cierva'' Fellowship.

\end{document}